%% file: main.tex
\documentclass{jfm_modified_2021}
\usepackage{graphicx}
\usepackage{amsmath, amssymb}
\usepackage[inactive,tightpage]{preview}
\usepackage{import}
\usepackage{microtype}
\usepackage[normalem]{ulem}
\usepackage{mathtools,bm}
\usepackage{cases}
\usepackage{esvect}
\usepackage{enumitem}

\pdfmapfile{+rsfso.map}
\DeclareSymbolFont{rsfso}{U}{rsfso}{m}{n}
\DeclareSymbolFontAlphabet{\mathscr}{rsfso}

\usepackage{pdflscape}
\usepackage{afterpage}
\usepackage{flafter}
\usepackage{rotating}
\usepackage{fancyhdr}
\fancypagestyle{lscape}{%
\fancyhf{} 
\fancyfoot[LE]{}
\fancyfoot[LO] {}
 
}

\usepackage{bm}
\usepackage{microtype}

\usepackage{hyperref}
\usepackage[usenames, dvipsnames]{xcolor}
\hypersetup{colorlinks=true,linkcolor=MidnightBlue,citecolor=MidnightBlue}
\usepackage{natbib}
\usepackage{array}
\usepackage{booktabs}
\usepackage{multirow}
\usepackage{siunitx}
\usepackage{mathtools}
\usepackage{cleveref}

\usepackage{caption}
\usepackage{subcaption}
\usepackage{tabularx}
\newcolumntype{Y}{>{\centering\arraybackslash}X}

\usepackage{tikz}
\usepackage{pgfplots}
\usepackage{pgfplotstable}
\usetikzlibrary{backgrounds, calc, patterns.meta, shadows, arrows, patterns}
\usepackage[outline]{contour} 
\contourlength{0.15em}

\usetikzlibrary{external}

\usepackage{grffile}
\pgfplotsset{compat=newest}
\usetikzlibrary{plotmarks}
\usetikzlibrary{arrows.meta}
\usepgfplotslibrary{patchplots}
\usepgfplotslibrary{fillbetween}

\usepackage{listings}

\shorttitle{A calibration-free QMED model}
\shortauthor{Morawiecki and Trinh}

\title{A calibration-free physicality-based model for predicting peak river flows}

\author{Piotr Morawiecki\corresp{\email{pwm27@bath.ac.uk}}
 \and Philippe Trinh\corresp{\email{hppt20@bath.ac.uk}}}

\affiliation{
    Department of Mathematical Sciences, University of Bath, Bath BA2 7AY, UK
}

\pubyear{XXXX}
\volume{YYY}
\pagerange{xxx--xxx}
\date{\today~[Draft]}

\newcommand{\tsat}[0]{t_\mathrm{crit}}
\newcommand{\qsat}[0]{q_\mathrm{crit}}

\begin{document}

\maketitle

\begin{abstract}
Many simple hydrologic models are based on parametric statistical relations between the river flow and catchment properties such as its area, precipitation rates, soil properties, etc., fitted to the available data. The main objective of this work is to explain how these statistical relations emerge from the physical laws governing surface and subsurface flow at a catchment scale.
The main achievement of this work is the derivation of an analytic formula for predicting peak monthly and annual river flows. It does not require any parameter calibration, but requires a measurement or estimation of the mean flow at the given catchment's outlet. We found that this model 1)~has a simple physical interpretation, 2)~provides more precise estimates than the median maximum annual flow (QMED) estimation method from the Flood Estimation Handbook (FEH), commonly used to estimate flood risk in the ungauged catchments in the UK, and 3)~is highly accurate for all types of catchments, including the small catchments, for which the standard FEH method is the least accurate.
\end{abstract}



\section{Introduction}

\noindent Predicting peak river flows is one of the key challenges in hydrology, with important applications, \emph{e.g.} in flood estimation. Over the years, many models have been developed to address this problem \citep{peel2020historical}. In the UK, the Flood Estimation Handbook (FEH) recommends using an estimation method by \cite{kjeldsen2008improving}, which provides the following formula for the median of the annual peak flow (QMED):
\begin{equation}
    \label{eq:feh_qmed}
    \text{QMED} = 8.3062\:\text{AREA}^{0.8510}\:0.1536^\frac{1000}{\text{SAAR}}\:\text{FARL}^{3.4451}\:0.0460^{\text{BFIHOST}^2},
\end{equation}
where AREA, SAAR, FARL, and BFIHOST are catchment descriptors, discussed later in \cref{sec:statistical_methods}. Essentially, this is a regression formula with four covariates and five free parameters that were fitted to the data collected in gauged UK catchments. It is one of many statistical models that have been developed over the last few decades, each consisting of different covariates, regression formulas, and fitted parameters \citep{faulkner2012estimating}. They are an outcome of exploratory modelling, in which the model is formulated to match the observed data rather than understand the underlying mechanism for peak flow generation. The drawback of this approach is that such models are restricted by the available training data, and we no guarantee that they would perform equally well in the situations underrepresented in the training dataset \citep{klemevs1986operational,beven2014glue}.

In this work, we suggest an alternative approach that could facilitate the standard data-exploratory development of statistical models. Instead of formulating many different statistical models based on the available data, we postulate formulating them based on theoretically-justified assumptions characterising physical hydrologic models and reducing them to simple scaling laws, like the one captured by the QMED equation \eqref{eq:feh_qmed}. This way, we can build simple, well-theoretically justified benchmark models that, as we shall show, may better inform us on parameters impacting peak flows and possible limitations of other approaches.

The goal of this study is to present a proof of concept of this new approach to model development. The main result of this study is the derivation and assessment of a simple expression for the peak flow, $q$:
\begin{equation}
    \label{eq:our_qmed}
    q=\bar{q}+\left(r-\bar{r}\right)\frac{\bar{q}}{\bar{r}}\left(1 - \text{BFI}\right).
\end{equation}
This equation is based only on four covariates: (i) the mean river flow at the outlet $\bar{q}$, (ii) the mean precipitation rate $\bar{r}$, (iii) the peak precipitation rate $r$, with the same return period as the desired flow $q$, and (iv) the base flow index $\text{BFI}$, defined as the ratio between base flow and total river flow. Unlike \eqref{eq:feh_qmed}, this equation does not include any fitted parameters and thus does not require any training data or calibration.

\begin{figure}
    \centering
    \includegraphics{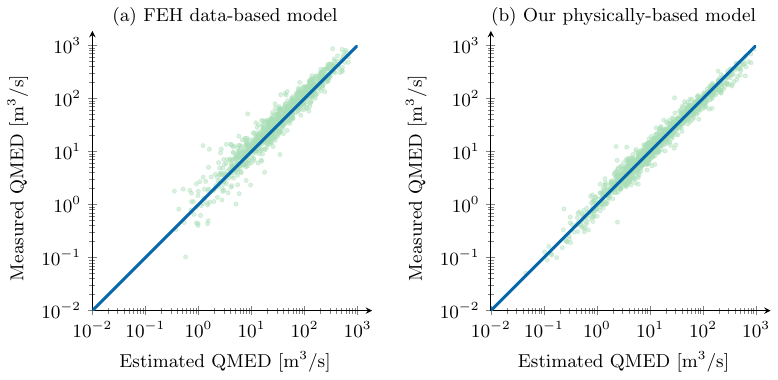}
    \caption{Comparison of the FEH data-based model \eqref{eq:feh_qmed} and our physically-based model \eqref{eq:our_qmed} for estimating the median of the annual peak flow (QMED). The QMED in the FEH method is calculated based on the AMAX dataset, while QMED in our method is calculated based on the Gauged Daily Flow dataset (see further discussion in \cref{sec:data_sources}).}
    \label{fig:model_comparison}
\end{figure}

As we shall show, this model can be used to provide unbiased predictions of monthly and annual peak flows. It seems to provide more precise predictions than flows estimated with the FEH method (see \cref{fig:model_comparison}). From the figure, notice how the FEH data-based model becomes increasingly inaccurate for small catchments with low QMED values, while our physically-based model is accurate over the entire domain. This is a crucial difference since one of the main applications of the QMED estimation method is to predict the flow in small ungauged sites, which is considered to be one of the key problems in catchment hydrology \citep{moore2007rainfall}.

Although \eqref{eq:our_qmed} cannot be directly applied to ungauged catchments with unknown mean flows, we can use the conclusions from this model to understand the limitations of the FEH method beyond what can be learned from statistical and data analysis alone. Therefore, following some authors (\emph{e.g.} \citealt{kirchner2006getting} and \citealt{wagener2010future}), we believe that a change of focus from purely data-oriented to more theoretically-justified models is essential to develop more robust catchment models to address this and many other problems of modern-day hydrology.

\section{Literature overview}


\subsection{On hydrologic models and their limitations}

Many models have been formulated over the years to address the aforementioned problem of predicting peak river flows. As noted by \emph{e.g.} \citet{moore2007rainfall} and \citet{sitterson2018overview}, these models can be divided into three categories: (i) physically-based models; (ii) conceptual models; and (iii) statistical models. Physical models are based on the fundamental laws of fluid dynamics to describe the surface and subsurface flow at a catchment scale. Conceptual models use physical concepts, such as the separation of surface and subsurface flows, but describe them with simpler mathematical models, requiring less data and computational resources. Finally, we have statistical models, which are based on finding a statistical relation between catchment properties and observed peak flows using, for example, regression methods.

Typically, these models are developed independently, with only a few intercomparison studies conducted. An example of such a study is the work by \cite{calver2009comparative}, which compared flood frequency predictions between statistical models (such as flood peak and event-based methods of the Flood Estimation Handbook) and conceptual models (such as the five-parameter Probability-Distributed Model and the Time-Area Topographic Extension model).

Statistical models, unlike physically-based and conceptual models to some extent, lack a foundation in understanding the underlying physical principles (except for the choice of covariates). Consequently, they heavily rely on available data and are strictly limited by the training data at hand. For example, there is no guarantee that a statistical model trained to predict flood frequencies based on historical data will perform well in the future, \emph{e.g.} due to potential changes in the hydrologic cycle caused by climate change \citep{wagener2010future}.

To address this limitation, \cite{formetta2018estimating} proposed using a conceptual Grid-to-Grid model to predict the median annual maximum flow. However, the same limitation may apply, to some extent, to conceptual models as their calibration is also restricted by the available training data. If a conceptual model is trained using data collected over the last 10 years, its performance in predicting flows with a return period of 100 years, resulting from extreme rainfall, is uncertain. The issue of general model transposability has been highlighted by \cite{klemevs1986operational} and more recently by \cite{beven2019make}, who has made efforts to develop more rigorous model testing methodologies \citep{beven2018hypothesis}.

An alternative approach is to use physically-based models as benchmarks for data-driven approaches. However, this area has been relatively unexplored due to the complexity of physical models and their dependence on detailed spatial data, which is often unavailable for most catchments \citep{moore2007rainfall}. Nonetheless, simplified catchment scenarios, such as those proposed in intercomparison numerical studies by \cite{sulis2010comparison} and \cite{maxwell2014surface}, can be utilised. In our previous work, we developed and studied a simple physical benchmark scenario based on standard governing equations for coupled surface-subsurface flow \citep{paper1, paper2}, and even provided an analytic solution for river flow formation in overland-dominated catchments \citep{paper3}.

In our later study \citep{rr1, rr2}, we demonstrated the utility of this benchmark model by assessing two conceptual models: the Probability-Distributed Model (PDM) by \cite{lamb1999calibration} and the aforementioned Grid-to-Grid model by \cite{bell2007development}. Understanding the similarities and differences between the physical and conceptual models allowed us to identify sources of inaccuracies and the limits of applicability of these models. We conducted this analysis based on a simple catchment model, but an open question remains regarding how closely the predictions of these simple benchmark models align with the properties of much more complex real-world systems.

Therefore, in this paper, we extend these studies to compare physical and statistical models. We aim to address two main questions:
\begin{enumerate}[label={(\roman*)},leftmargin=*, align = left, labelsep=\parindent, topsep=3pt, itemsep=2pt,itemindent=0pt]
    \item Does a physical benchmark model predict the same statistical relationships between catchment properties as those observed in nature?
    \item Can physical benchmarks be used to gain a better understanding of the physical foundations behind statistical models?
\end{enumerate}

This demonstrates a new possible approach in the development of hydrologic models. Instead of constructing statistical models based on exploratory studies of available data, one can use a physical model based on clear theoretical foundations. Models constructed in this way have a well-defined structure and well-understood limits of applicability, making them potentially more reliable when applied to situations not included or underrepresented in the training data (\emph{e.g.} when applied to small ungauged catchments).

\subsection{On the statistical peak-flow estimation methods}
\label{sec:statistical_methods}

Historically, statistical methods started being developed long before the rise in popularity of conceptual and physical models during the digital revolution. They are still popular among practitioners since they are easy to use and do not require any computing power for practical application. Statistical models, fitted to available data from gauged catchments, are commonly used to predict flow properties at ungauged sites. Examples of such models include the Institute of Hydrology mean annual maximum flow (QBAR) estimation method \citep{marshall1994flood}, the median annual maximum flow (QMED) method from the Flood Estimation Handbook (FEH) \citep{reed1999flood}, and the Revitalised Flood Hydrograph (ReFH) method by \cite{kjeldsen2008improving}. For a comprehensive review of these and other statistical models, refer to \cite{faulkner2012estimating} or \cite{fleig2013flood}.

In this report, we study the case of the FEH method, which uses a regression equation to estimate QMED, defined as the median of the annual maximum flow series. Originally formulated by \cite{reed1999flood} for rural catchments in the UK and updated by \cite{kjeldsen2008improving}, it uses four covariates: $\text{AREA}$, the catchment drainage area [$\mathrm{km^2}$]; $\text{SAAR}$, the standard average annual rainfall [mm/year]; $\text{FARL}$, the index of flood attenuation due to reservoirs and lakes; and $\text{BFIHOST}$, the base flow index derived from the HOST classification by \cite{boorman1995hydrology}. FARL is equal to 1 if no reservoirs or lakes are present. The base flow index is defined as the ratio of long-term base flow to the total river flow, and is close to 1 for groundwater-dominated catchments and close to 0 for overland-dominated catchments. The regression formula fitted to 602 gauged rural catchments in the UK yields:
\begin{equation}
    \label{eq:qmed}
    \text{QMED} = 8.3062\:\text{AREA}^{0.8510}\:0.1536^\frac{1000}{\text{SAAR}}\:\text{FARL}^{3.4451}\:0.0460^{\text{BFIHOST}^2}.
\end{equation}

\begin{figure}
    \centering
    \includegraphics{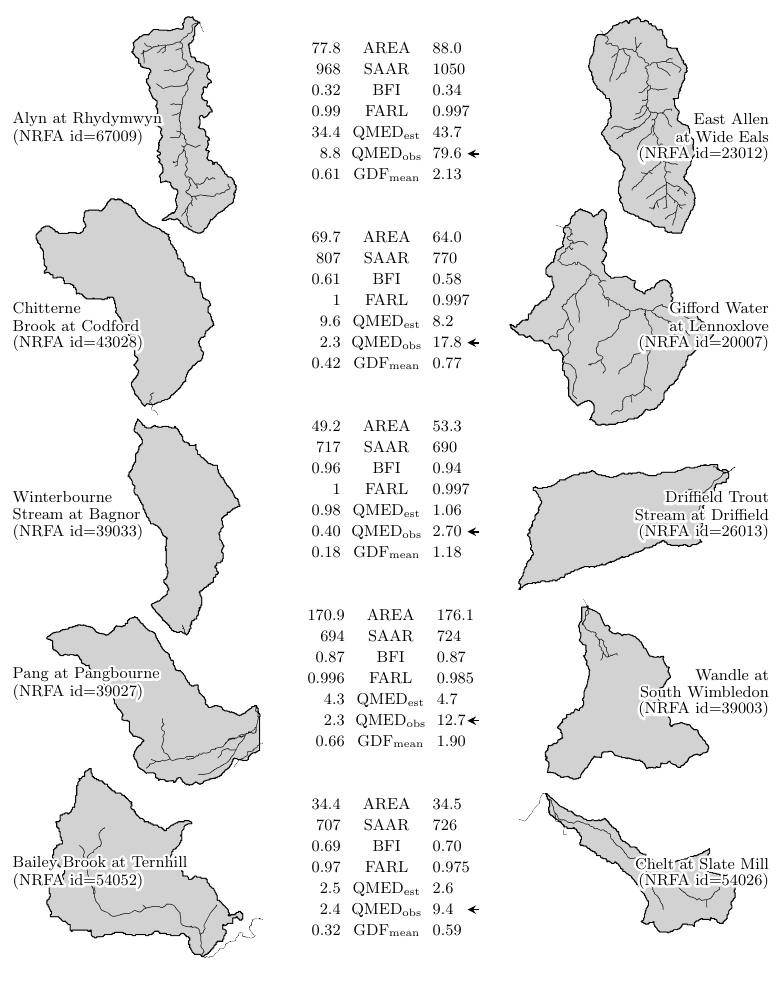}
    \caption{Examples of pairs of catchments with almost the same FEH catchment descriptors, but significantly different observed mean and annual peak river flows.}
    \label{fig:similar_catchments}
\end{figure}

Despite providing accurate predictions for most of the training data, predictions can be highly inaccurate for some sites \citep{asadullah2018estimating}. The most problematic cases are small catchments due to (1) fewer flow records being available for small catchments as their flow is not gauged and (2) the larger influence of local features on floods in such catchments. As an example, consider pairs of catchments presented in \cref{fig:similar_catchments}. Each pair has almost the same catchment descriptors used in \cref{eq:qmed}, resulting in similar estimated QMED values. However, the actual measured QMED values are significantly different. This indicates that in some catchments, the fitted formula should not be used, but so far, the limits of its applicability are not specified.

A similar difference can be observed in the mean river flow in these catchments, which demonstrates that there are significant factors affecting the mean flow that are not represented by the FEH catchment descriptors. Information provided in the National River Flow Archive (\url{https://nrfa.ceh.ac.uk/}) on the flow regime in the selected catchments provides examples of such factors:

\begin{enumerate}[label={(\roman*)},leftmargin=*, align = left, labelsep=\parindent, topsep=3pt, itemsep=2pt,itemindent=0pt]
    \item\;In the first pair, the Alyn catchment experiences substantial losses of river flow through percolation to mine discharge tunnels, which may explain the much lower mean flow compared to a similar East Allen catchment.
    \item\;In the second pair, there is a clear visual difference between the catchments. Chitterne Brook has river channels and hardly any overland flow, so the water flow (and total evapotranspiration) is expected to be far greater than in the Grifford Water catchment, characterised by a well-developed drainage network.
    \item\;In the third pair, the runoff in the Winterbourne Stream catchment is reduced by groundwater abstractions, \emph{e.g.} through the West Berkshire Groundwater Scheme.
    \item\;In the last two pairs, despite the catchments having almost the same FEH descriptors, their land cover differs significantly. On the left, we have two rural catchments, while on the right, we have urban catchments. In the case of rural catchments, the resulting runoff is much lower than for urban catchments due to factors such as evapotranspiration from vegetation.
\end{enumerate} 

Neither of these effects is represented by the FEH catchment descriptors, which, as we argue further in this paper, is the main source of the inaccuracy of the FEH model \eqref{eq:qmed}.

\subsection{Peak-flow generation according to the physical benchmark}
\label{sec:peak_flow_generation}

Both statistical models and physical models have proven to be successful in predicting river flows. Therefore, we expect that similar scaling laws to the ones postulated in the statistical regression models should also correspond to the scaling laws characterising the predictions of the physical model.

In our previous study \cite{paper1, paper2, paper3}, we applied a physical model to a simple benchmark scenario. The main idea was to produce a minimal model that is complex enough to include all the key physical mechanisms responsible for flow generation, but simple enough to fully understand its structure and investigate its properties with semi-analytic tools.

The benchmark geometry that we considered is a three-dimensional V-shaped catchment with uniform soil and surface hydraulic properties. This can be reduced to a two-dimensional hillslope model or even to a one-dimensional model for low-productivity catchments (\emph{i.e.} with subsurface flow limited to a thin layer of porous soil). In all these geometries, we investigated the shape of the hydrograph observed during a single rainfall event.

\begin{figure}
    \centering
    \includegraphics{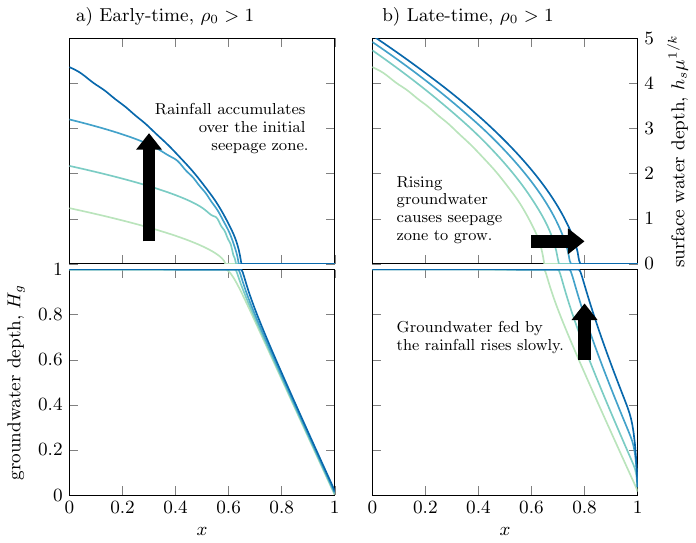}
    \caption{Schematic presenting early- and late-time dynamics for a catchment with an initial seepage zone (corresponding to $\rho_0>1$).}
    \label{fig:early_late_time_behaviour}
\end{figure}

As our in-depth analysis of low-productivity catchments (\emph{i.e.} with subsurface flow limited to a thin layer of porous soil) has demonstrated \citep{paper3}, the hydrograph has a qualitatively different shape depending on a dimensionless parameter:
\begin{equation}
    \rho_0=\frac{r_0 L_x}{K_s S_x L_z},
\end{equation}
interpreted as the ratio between the mean precipitation and groundwater flow. If $\rho_0>1$, the soil up to a certain distance from the river is fully saturated (which we refer to as the \textit{seepage zone}). The fraction of the hillslope covered by the seepage zone is $a_0=1-\frac{1}{\rho_0}$, see \cref{fig:early_late_time_behaviour}.

\begin{figure}
    \centering
    \includegraphics{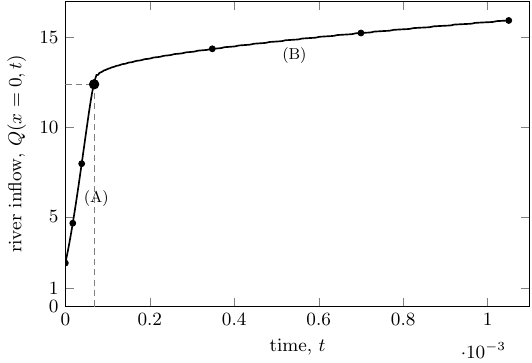}
    \caption{Schematic representation of a hydrograph obtained for a catchment with an initial seepage zone. Points represent times for which the profiles are shown in \cref{fig:early_late_time_behaviour}, while letters (A) and (B) refer to corresponding phases from that figure.}
    \label{fig:example_hydrographs}
\end{figure}

If the seepage zone exists, two distinct timescales are observed on a storm hydrograph (see \cref{fig:example_hydrographs}). During a storm, the water accumulates over the seepage zone and quickly reaches the river. This causes the river flow to experience a fast early-time growth from $q_0=r_0 L_x$ to the critical flow $\qsat$ given by
\begin{equation}
    \label{eq:qsat_dimensional}
    \qsat = \underbrace{K_s S_x L_z}_{\text{groundwater flow}} + \underbrace{r L_x \left(1 - \frac{K_s S_x L_z}{r_0 L_x}\right)}_{\text{overland flow}},
\end{equation}
reached at the critical time, $\tsat$, as shown in \cref{fig:example_hydrographs}. The critical flow expressed above is the river inflow per unit length of the river in [$\mathrm{m/s^2}$].

For $t>\tsat$, we observe a much slower rise of river flow caused by the slow rise of the groundwater table, leading to the seepage zone slowly growing and increasing the area over which overland flow is formed. The clear distinction between the early- and late-time phases is caused by a huge difference in timescales characterising overland and groundwater flow (typically different by three orders of magnitude).

The presented model was developed under strict assumptions. Firstly, the geometry was limited to a simple V-shaped catchment with homogeneous surface and subsurface properties, and then reduced to a one-dimensional hillslope model. Other assumptions include constant and uniform precipitation rate during rainfall and subsurface flow being limited to a thin layer of soil. However, as we argue in the next section, the existence of the aforementioned critical flow can be generalised to complex drainage networks.

\section{Methods}
\label{sec:methods}

Our simple physical benchmark model allows us to find a benchmark hydrograph during a single rainfall event. In this section, based on the conclusions obtained from these benchmark hydrographs, we demonstrate how one can estimate annual peak river flows, which are essential for flood estimation. Firstly, we show how peak flows can be estimated in the 1D hillslope geometry, and then we generalise it to more complex geometries so that its predictions can be compared with the peak flows observed in real-world systems.

\subsection{Discussion based on 1D hillslope geometry}

Typically, the contribution of the critical flow to the total flow of the river is greater than the additional late-time flow rise. We demonstrated this in Part 3 of our previous work by performing a sensitivity analysis over a wide range of parameter values characterising real-world catchments. For example, under typical conditions, the critical flow contributes to approximately 75\% of the peak flow reached by extreme 24-hour rainfall. Therefore, to understand how the peak flow scales with catchment parameters, we should investigate their impact on critical flow in the first place.

Note that the total river inflow $Q=\qsat L_y$ at the critical time can be expressed as:
\begin{equation}
    \label{eq:Qpeak_formula_1D_model}
    Q = K_s S_x L_z L_y + r L_x L_y \left(1 - \frac{K_s S_x L_z}{r_0 L_x}\right) = r_0\cdot\mathrm{AREA}\cdot\mathrm{BFI} + r\cdot\mathrm{AREA}\cdot(1 - \mathrm{BFI}),
\end{equation}
where $\mathrm{AREA}=L_xL_y$ is the catchment area and $\text{BFI}=\frac{K_s S_x L_z}{r_0 L_x}$ is the ratio between the mean groundwater flow ($K_s S_x L_z$) and mean total river flow ($r_0 L_x$). Since, in the absence of rainfall, the overland flow quickly drains, the remaining groundwater flow plays the role of long-term base flow. Therefore, this fraction is equivalent to the Base Flow Index used in the statistical models.

To summarise, we can express the dominant part of the total river inflow using only four descriptors: $r$, $r_0$, $\mathrm{AREA}$, and $\mathrm{BFI}$. Analogous descriptors are currently used in the FEH statistical flood, except for two differences. Firstly, in addition to the mean precipitation rate $r_0$ corresponding to Standardised Annual Average Rainfall (SAAR) FEH descriptor, we also have the peak precipitation rate that corresponds to the simulated rainfall. We will comment on it in \cref{sec:data_sources}. Secondly, an additional parameter FARL describing the attenuation influence of open water bodies on the peak flows. However, our benchmark scenario does not include any reservoirs or lakes, so it corresponds to a catchment with $\mathrm{FARL}=1$.

\subsection{Generalisation for any catchment geometry}

In this section, we use observations from the previous paragraph to show that the result~\eqref{eq:Qpeak_formula_1D_model} can, under certain assumptions, be applied to real-world catchments described by more complex geometries and complex aquifer structures.

\begin{figure}
    \centering
    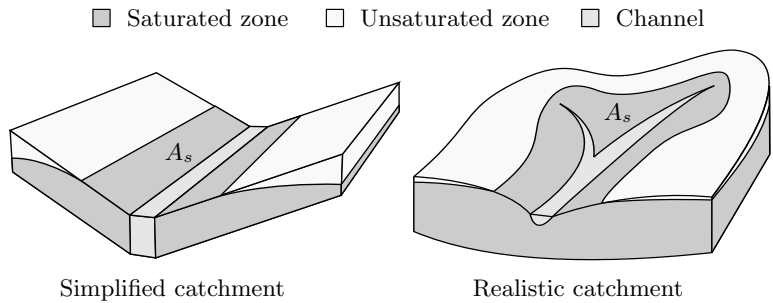
    \caption{Illustration of seepage zone in simplified and realistic catchment.}
    \label{fig:saturated_zone}
\end{figure}

Let us consider a catchment of area $A$ with an arbitrary geometry, for which some part of the soil around the river bed is fully saturated. In other words, there exists a seepage zone, which forms when overland water occurs during rainfall (see \cref{fig:saturated_zone}). Then the groundwater flow $Q_g$ can be assumed to be constant.

The critical flow is given by precipitation accumulated over the seepage zone $A_s$:
\begin{equation}
    \label{eq:critical_flow_general}
    q-\bar{q}=\left(r-\bar{r}\right)A_s.
\end{equation}
Here, $q$ and $\bar{q}$ are the peak and mean flow ($m^3/s$), while $r$ and $\bar{r}$ are the peak and mean precipitation rate ($m/s$). Now we want to estimate the size of the seepage zone using the known value of the base flow index.

The mean flow is composed of a constant groundwater flow $Q_g$ [$m^3/s$] and the mean precipitation accumulated over the seepage zone:
\begin{equation}
    \bar{q} = \bar{r} A_s + Q_g.
\end{equation}
The base flow index (BFI) is defined as the ratio between the groundwater flow and the total flow:
\begin{equation}
    \label{eq:bfi_equation}
    \text{BFI} = \frac{Q_g}{\bar{q}} = \frac{\bar{q} - \bar{r} A_s}{\bar{q}} = 1 - \frac{\bar{r}}{\bar{q}}A_s.
\end{equation}
Therefore, from~\eqref{eq:bfi_equation}, we derive an expression for the size of the seepage zone:
\begin{equation}
    A_s = \frac{\bar{q}}{\bar{r}}\left(1 - \text{BFI}\right).
\end{equation}
By substituting it into~\eqref{eq:critical_flow_general}, we obtain an expression for the peak flow in terms of catchment parameters:
\begin{equation}
    \label{eq:QMED_new}
    q=\bar{q}+\left(r-\bar{r}\right)\frac{\bar{q}}{\bar{r}}\left(1 - \text{BFI}\right).
\end{equation}
Alternatively, one can express the mean flow $\bar{q}$ as the difference between the mean precipitation $\bar{r}$ and the mean evapotranspiration $\bar{E}$ over the entire catchment area, \emph{i.e.} $\bar{q}=\left(\bar{r}-\bar{E}\right)A$. Substituting it into~\eqref{eq:QMED_new} gives:
\begin{equation}
    \label{eq:QMED_new_E}
    \begin{split}
        q&=\bar{r}\left(1-\frac{\bar{E}}{\bar{r}}\right)A+\left(r-\bar{r}\right)\left(1-\frac{\bar{E}}{\bar{r}}\right)A\left(1 - \text{BFI}\right)\\
        &=\left(1-\frac{\bar{E}}{\bar{r}}\right)A\bigg(r\left(1-\text{BFI}\right)+\bar{r}\;\text{BFI}\bigg)
    \end{split}
\end{equation}
Note that in the case of no evaporation, \eqref{eq:QMED_new_E} is reduced to\eqref{eq:Qpeak_formula_1D_model}. However, as the mean evaporation increases, the peak flow theoretically decreases, eventually reaching 0 as $\bar{E}\rightarrow\bar{r}$.

Equations~\eqref{eq:QMED_new} (or~\ref{eq:QMED_new_E}) allow us to find the peak flow, given the mean and peak precipitation rates ($\bar{r}$ and $r$), the base flow index (BFI), the catchment's area, and the mean flow $\bar{q}$ (or mean evapotranspiration $\bar{E}$) for a given catchment. Depending on which statistical property of the flow we want to estimate, we should take appropriate values for the remaining parameters. For example, to find the median of the maximum flow in January, we need to take $r_0$ and $q_0$ equal to the mean precipitation in January, and $r$ equal to the median of the maximum precipitation in January. In the further course of this study, we will verify how accurate model~\eqref{eq:QMED_new} is for different seasons when applied to the UK catchments.

\subsection{Data sources and preprocessing}
\label{sec:data_sources}

To assess the accuracy of the presented model, we used data from the National River Flow Archive (NRFA), which is a comprehensive database of UK river flows mandated by the UK government (Defra) and the devolved administrations of Northern Ireland, Scotland, and Wales \citep{hannaford2004development}. This database consists of data from over 1602 gauged catchments, some of which are already closed, while others are still operating.

In this study, we utilised three datasets from the NRFA. Firstly, we used the Catchment Information dataset, which includes general information about the catchments, such as parameters from the Flood Estimation Handbook (FEH) QMED estimation method (AREA, SAAR, FARL, BFIHOST) and the base flow index (BFI) calculated directly from daily flow data. For model comparison, we used the direct measurement of BFI, as using BFIHOST introduces additional uncertainty to both models, \eqref{eq:qmed} and \eqref{eq:QMED_new}, which can overshadow their true accuracy. However, when dealing with ungauged catchments, daily flow data is not available, so we will report the accuracy of both models when BFIHOST is applied as well.

Secondly, to extract $\bar{r}$ and $r$, we used the Catchment Rainfall Data, which includes cumulative daily rainfall (CDR) data in millimetres. Finally, to extract $\bar{q}$ and $q$, we used the Daily River Flow Data, which includes gauged daily flow (GDF) measured in cubic meters per second ($\mathrm{m^3/s}$). In both of these datasets, we extracted the mean and maximum values of CDR and GDF for each month and year included in the database. We estimated $r$ and $q$ by taking the median of the maximum values of the given quantity for each month, while $\bar{r}$ and $\bar{q}$ were estimated by taking the median of the mean values of CDR and GDF for each month, respectively. This approach is equivalent to taking the peak flow $q$ and peak rainfall $r$ with a two-year return period. Similarly, one can estimate peak flows $q$ with different return periods by extracting the peak precipitation $r$ data with the same return period. However, this study focuses on the estimation of QMED, \emph{i.e.} the flow with a two-year return period.

In total, we processed data from 1538 UK catchments, as some catchments had incomplete or missing datasets. The scripts used to extract, process, and post-process the data were written in the R programming language and can be found in our GitHub repository \citep{github_qmed}. We utilised the NRFA API to automate the data processing.

\section{Results}

\subsection{Monthly peak flow estimation}
\label{sec:monthly_peak_flow_est}

\begin{figure}
    \centering
    \includegraphics{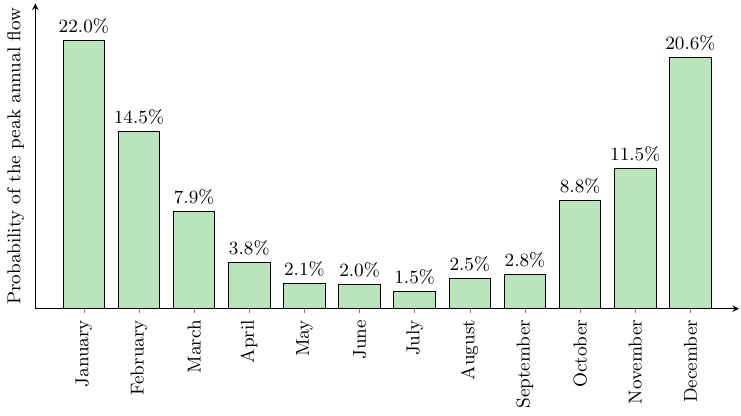}
    \caption{Distribution of months during which the peak daily flows occur. The probability was calculated based on the GDF dataset for all catchments in the NRFA, restricted to years with complete data for all months.}
    \label{fig:month_distribution}
\end{figure}

As mentioned previously, the value of catchment parameters appearing in \eqref{eq:QMED_new} depends on the season. According to \eqref{eq:QMED_new_E}, the highest flows occur when mean and peak precipitation are at their highest and evapotranspiration is at its lowest, which in the case of the UK climate mostly happens in winter. As shown in \cref{fig:month_distribution}, the majority of peak flows occur in January ($22\%$), followed by December ($20.6\%$) and February ($14.5\%$). Therefore, we will focus on investigating the peak flows in January and perform standard statistical diagnostics. The precision of the model for other months is briefly presented in \cref{app:monthly_summary}.

\begin{figure}
    \centering
    \includegraphics{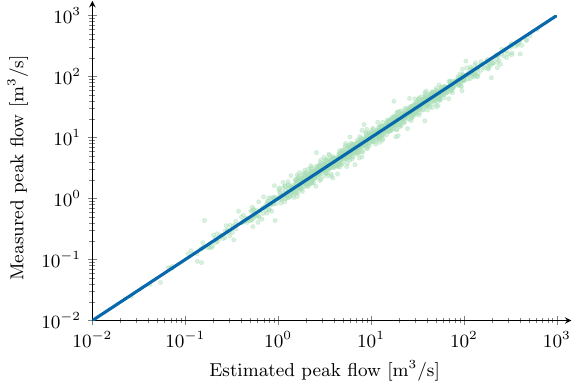}
    \caption{Comparison of the median of maximum daily flow in January estimated using~\eqref{eq:QMED_new} and measured based on gauged daily flow for 1538 UK gauged catchments.}
    \label{fig:our_model_fit}
\end{figure}

Fig.~\ref{fig:our_model_fit} illustrates the relationship between the median of maximum peak flow in January estimated by equation~\eqref{eq:QMED_new} and the actual observed value of $q$ at the gauging stations. The red line represents equal estimated and actual $q$ values. For each point, we calculate log residuals, given by:
\begin{equation}
    \label{eq:residuals}
    e_i = \log_{10}\left(q^{\text{real}}_i\right) - \log_{10}\left(q^{\text{est}}_i\right) = \log_{10}\left(q^{\text{real}}_i\right) - \log_{10}\left(\bar{q} + \left(r_i-\bar{r}_i\right)\frac{\bar{q}_i}{\bar{r}_i}\left(1 - \text{BFI}_i\right)\right)
\end{equation}
We use logarithm to estimate the size of the relative error rather than the absolute error, so that both small and large catchments have the same contribution to the mean square error. The base 10 is chosen for easier interpretation of the calculated residuals, where $e = \pm 1$ represents an estimated value lower/higher than the real value by exactly one order of magnitude.

For the January data, the mean value $\text{MEAN}(e)=-0.014$ is negative, indicating that the model slightly overestimates the actual peak flows. However, since this bias is much smaller than the standard deviation of residuals, $\text{SD}(R)=0.08$, we conclude that the bias is very small. A similar bias can be observed in other months (see \cref{tab:monthly_summary}), with the greatest bias in the summer months and the lowest in winter. However, as argued earlier, the winter months are responsible for over half of the observed annual peak flows and therefore are the most important in predicting annual peak flows.

To assess the quality of fit, we used the coefficient of determination $R^2$:
\begin{equation}
    R^2 = 1 - \frac{\sum_{i=1}^n \left(\log_{10} q^{\text{real}}_i - \log_{10} q^{\text{est}}_i\right)^2}{\sum_{i=1}^n \left(\log_{10} q^{\text{real}}_i - \sum_{j=1}^n \log_{10} q^{\text{real}}_j\right)^2},
\end{equation}
which is defined as the proportion of the variation in the dependent variable that is predictable from the independent variables. For the peak flow data from January, $R^2=0.987$, indicating that our simple model \eqref{eq:QMED_new} explains approximately $98.7\%$ of the variation in the peak flow dataset.

\begin{figure}
    \centering
    \includegraphics{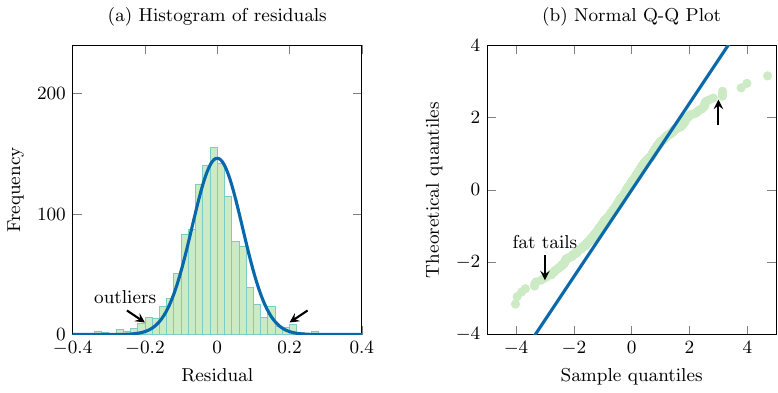}
    \caption{(A) Distribution of residuals from \cref{fig:our_model_fit}. A fitted normal distribution is represented by the solid line. (B) Q-Q Plot to examine whether the residuals follow a normal distribution. Deviation of points from a straight line indicates the occurrence of fat tails on both the positive and negative sides of the distribution.}
    \label{fig:our_model_res_distribution}
\end{figure}

The observed peak flow values are distributed around the estimated values, and the residuals follow an approximately normal symmetric distribution, as shown in \cref{fig:our_model_res_distribution}a. However, the far ends of the Q-Q plot in \cref{fig:our_model_res_distribution}b suggest the presence of heavy tails in the distribution, indicating the occurrence of outliers where the flow is significantly over- or underestimated.

\begin{figure}
    \centering
    \includegraphics{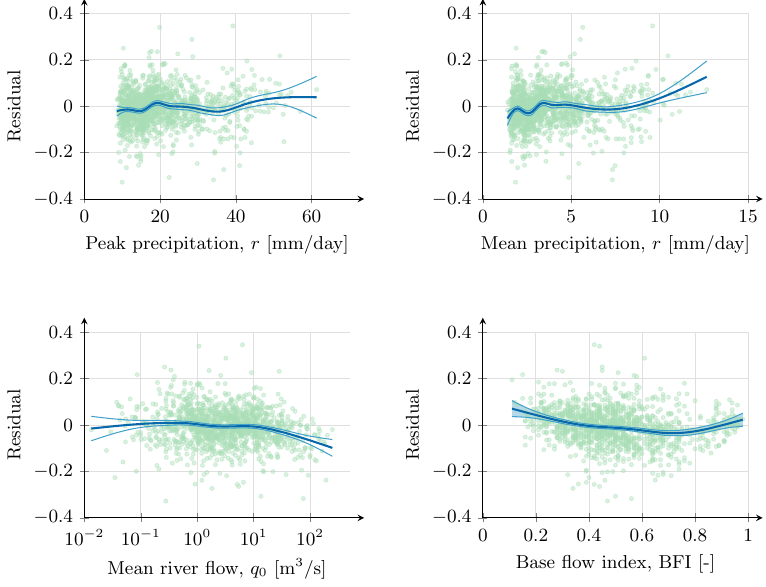}
    \caption{Dependence of the residuals on each of the four model covariates. The blue line was fitted using a generalised additive model (GAM).}
    \label{fig:our_model_residuals}
\end{figure}

Most importantly, we do not observe a significant dependence of the size of residuals on each of the four parameters, as illustrated in \cref{fig:our_model_residuals}. In regions where the mean residual deviates from the $R=0$ line, we typically have too few data points to draw any significant conclusions. It is worth noting that the residuals appear to have higher variance for catchments with low mean river flow $q_0$, but these data points correspond to small catchments where the river flow is gauged in very few cases. Therefore, we can expect a higher variance when applying this model to small ungauged catchments.

Based on these results, we conclude that model \eqref{eq:QMED_new} allows for precise and accurate prediction of peak flows in catchments with a significant seepage zone ($\text{BFI}<0.9$).

\subsection{Estimating the median annual maximum flow}
\label{sec:QMED_our_model}

Having peak flow estimates for each month of the year, we can estimate the maximum annual flow by finding the expected value of the maximum flow over all 12 months:
\begin{equation}
    \label{eq:QMED_def}
    \mathrm{QMED} = \mathbb{E}\left[\max_{m=1,\ldots, 12}\left(q^m\right)\right].
\end{equation}
Here, the upper index represents the month (1 - January, 2 - February, etc.), and $q_m$ is a random variable representing the peak flow in a given month. Unfortunately, finding the expected value presented above requires not only knowing the expected value of $q^m$ but also its full probability distribution. This is because the following inequality holds:
\begin{equation}
    \label{eq:QMED_inequality}
    \mathbb{E}\left[\max_{m=1,\ldots, 12}\left(q^m\right)\right] \leq \max_{m=1,\ldots, 12}\bigg[\mathbb{E}\left(q^m\right)\bigg],
\end{equation}
where $q$ is the estimate given by~\eqref{eq:QMED_new}. It is possible to estimate the probability distribution for $q^m$ for each month for gauged catchments with a sufficiently long GDF database, but 1) this procedure cannot be applied for ungauged sites for which the QMED method is applied, and 2) it would involve numerical estimation of the expected value in~\eqref{eq:QMED_def}, taking away the desired simplicity behind statistical methods. Instead, we estimate $\mathbb{E}\left(q^m\right)$ with our model~\eqref{eq:QMED_new}, additionally taking into account that it introduces an additional bias:
\begin{equation}
    \label{eq:QMED_model}
    \mathrm{QMED}=\max_{m=1,\ldots, 12}\bigg[\mathbb{E}\left(q^m\right)\bigg]-\mathrm{BIAS}=\max_{m=1,\ldots, 12}q\left(q_0^m, r^m, r_0^m, \mathrm{BFI}^m\right)-\mathrm{BIAS}.
\end{equation}
Two remarks should be made here. Firstly, since only average BFI is available in NRFA, we use the same value for each month. Secondly, we test this model using Gauged Daily Flow data, so we are estimating the maximum annual daily flow, rather than the maximum annual recorded peak flow. The latter one is given by the so-called AMAX series and is used in the Flood Estimation Handbook to estimate the QMED flood index, which can significantly exceed the mean daily flow value.

\begin{figure}
    \centering
    \includegraphics{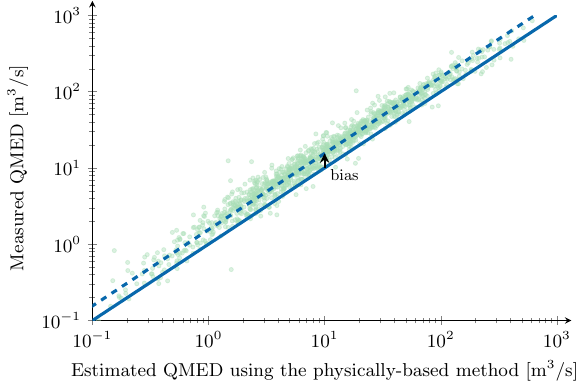}
    \caption{Comparison of the median of annual maximum daily flow, estimated using~\eqref{eq:QMED_new}, and measured based on gauged daily flow for 1538 UK gauged catchments. The solid line represents model~\eqref{eq:QMED_model} without included bias, and the dashed line represents the same model with $\mathrm{BIAS}=0.1957$.}
    \label{fig:QMED_our_model_fit}
\end{figure}

The median annual maximum flow from the GDF series, both derived directly from observations and from model~\eqref{eq:QMED_model}, is presented in \cref{fig:QMED_our_model_fit}. The red line represents the model predictions without bias. We can clearly see, as expected from inequality\eqref{eq:QMED_inequality}, that such a model overestimates the observed value of QMED. The mean value of the log residue is $0.1957$ and is approximately constant for all estimated QMED values. This is why we shall take a constant bias approximation $\mathrm{BIAS}=0.1957$. The model with the fitted bias is represented by the blue line.

After taking the bias into account, we can see that the~\eqref{eq:QMED_model} approximation is accurate, with the standard deviation of residual values equal to $\text{SD}(e)=0.119$ and the corresponding $R^2=0.971$. If instead of BFI derived from Gauged Daily Flow (which is unavailable for ungauged catchments), the BFIHOST estimate was used in~\eqref{eq:QMED_new}, the residual values would rise by over $25\%$ to $\text{SD}(e)=0.149$, corresponding to $R^2=0.955$. This shows that inaccuracies behind the BFIHOST classification introduce a significant error to the QMED estimation. We further discuss this in \cref{sec:discussion}.

\begin{figure}
    \centering
    \includegraphics{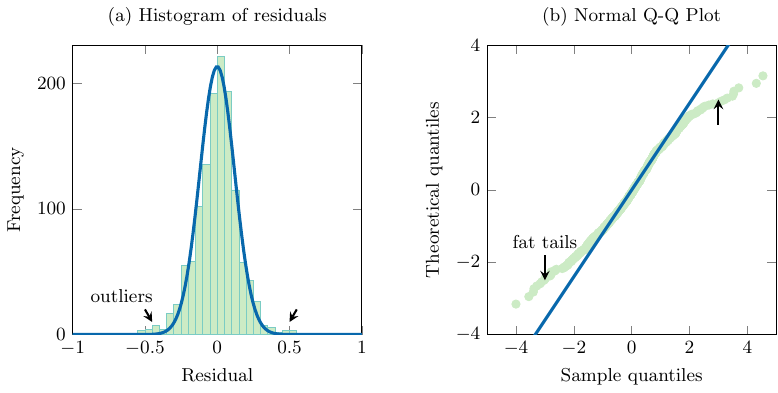}
    \caption{(A) Distribution of residuals from~\cref{fig:QMED_our_model_fit} with a fitted normal distribution. (B) Q-Q Plot comparing the distribution of residuals with a normal distribution.}
    \label{fig:QMED_model_res_distribution}
\end{figure}

As shown in \cref{fig:QMED_model_res_distribution}A, the residual values follow an approximately normal distribution. However, inspection of the Q-Q Plot in \cref{fig:QMED_model_res_distribution}B reveals fat tails on both ends of this distribution, similar to what we observed in \cref{fig:our_model_res_distribution} for the individual monthly peak flows.

\subsection{Comparison with QMED estimation method from FEH}

\begin{figure}
    \centering
    \includegraphics{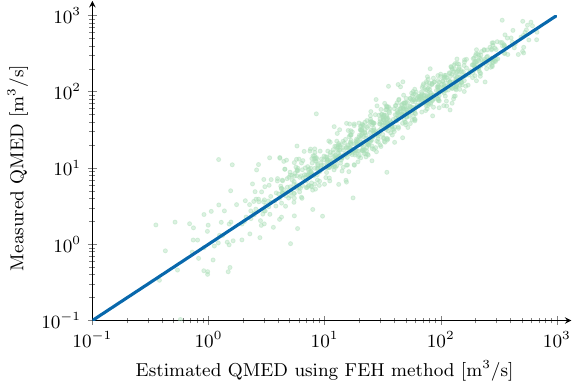}
    \caption{Comparison of the median annual maximum flow (QMED) estimated using our model and the QMED estimated based on the AMAX series for 939 catchments for which QMED values were available in the NRFA. The red line represents equal estimated and observed QMED values.}
    \label{fig:QMED_FEH_model_fit}
\end{figure}

In this section, we assess the accuracy of our model by comparing its predictions with the observed annual peak flows and predictions of the FEH QMED estimation method equation described in \cref{sec:statistical_methods}. Firstly, we estimate the value of QMED using our model and compare it with the measured value of QMED from the National River Flow Archive (NRFA). For the purpose of this comparison, we used the Base Flow Index derived from the Gauged Daily Flow dataset instead of the BFIHOST classification, in order to eliminate the impact of uncertainty in the HOST-based classification on the performance assessment of both QMED models. The comparison is presented in \cref{fig:QMED_FEH_model_fit}.

The standard deviation of the residual values ($\text{SD}(e)$) is equal to $0.174$ (or $0.186$ if BFIHOST was used), which is $46\%$ higher than the fit obtained with our model. Similarly, the $R^2$ value is significantly lower for the FEH method ($R^2 = 0.922$) compared to our model ($R^2 = 0.955$).

\begin{figure}
    \centering
    \includegraphics{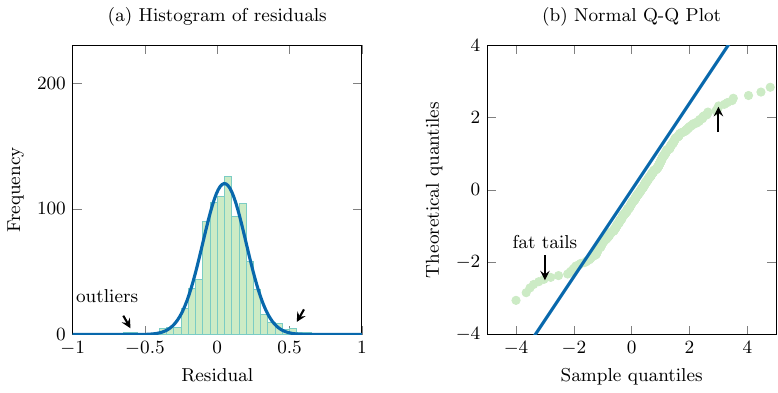}
    \caption{(A) Distribution of residuals from~\cref{fig:QMED_FEH_model_fit} with a fitted normal distribution. (B) Q-Q plot comparing the distribution of residuals with a normal distribution.}
    \label{fig:QMED_FEH_model_res_distribution}
\end{figure}

As in the case of the monthly model, the distribution of residuals in the FEH model is close to a normal distribution, as shown in \cref{fig:QMED_FEH_model_res_distribution}, with fat tails observed on both ends. However, based on the Q-Q plot comparison, the fat tails are thicker in the FEH model than in our model, indicating a higher number of extreme outliers.

As shown in~\cref{fig:QMED_FEH_model_fit}, most of the outliers appear in catchments with low QMED values. These catchments typically represent small catchments since QMED is approximately proportional to the catchment's area, or catchments with high BFI values dominated by groundwater flow. Our physically-based model provides a more accurate estimation in such situations, with precision comparable to that achieved for large catchments.

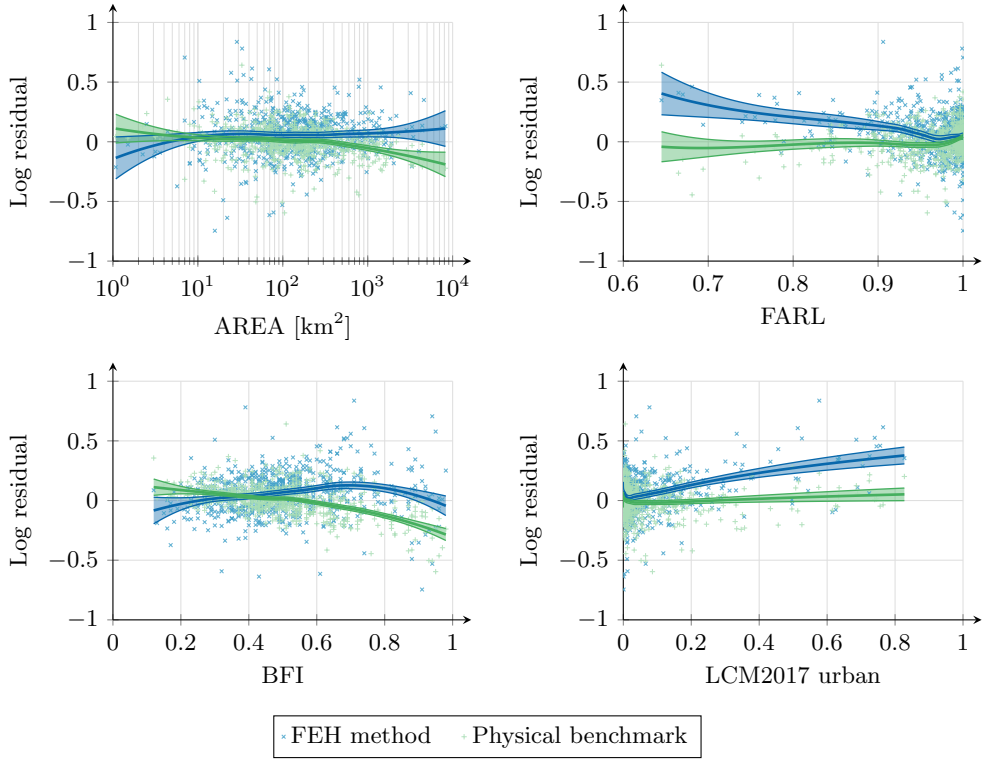
\begin{figure}
    \centering
    \input{TIKZ/res_vs_descriptors}
    \caption{Comparison of the residual distribution for FEH model and our physical benchmark for four different FEH descriptors: (a) catchment area, (b) Flood Attenuation by Reservoirs and Lakes, (c) Base Flow Index (derived from GDF series), and (d) urban coverage according to the Land Cover Map 2007. The shaded green and blue area represents the mean value of residuals for each model. It was fitted using a generalised additive model (GAM).}
    \label{fig:enter-label}
\end{figure}

\section{Discussion}
\label{sec:discussion}

We have demonstrated that a simple model based on physical principles can provide more accurate statistical predictions of annual peak flows (QMED) compared to using methods from the Flood Estimation Handbook (FEH), especially when applied to predict flow in small catchments. However, there are three major differences between these modelling approaches:
\begin{enumerate}[label={(\roman*)},leftmargin=*, align = left, labelsep=\parindent, topsep=3pt, itemsep=2pt,itemindent=0pt]
    \item\;The FEH QMED method predicts the median of the annual peak flows derived from AMAX series, while our physically-based method predicts the median of the annual peak daily flows derived from the Gauged Daily Flow dataset. One can adapt our method to the AMAX series, but it requires constructing (or estimating) the same series for rainfall for each month, \emph{i.e.} with the highest precipitation rate registered in each month.
    \item\;The physically-based method requires two parameters that are not used by the FEH method. The first parameter is the mean river flow $q_0$, which either has to be measured directly or potentially can be estimated based on water balance equations and appropriate evapotranspiration models. This parameter is crucial because it encodes the dependence between precipitation data and river flow data. The second parameter is the peak precipitation $r$ corresponding to the rainfall of the given return period (two years in the case of QMED estimation). This can be derived directly from the data or estimated statistically based on the known average annual rainfall (SAAR).
    \item\;QMED involves 5 parameters fitted to the data, while our method for monthly flow does not require any calibration, and our QMED method requires finding one parameter (BIAS) or calculating it with more complex probabilistic methods.
\end{enumerate}

Despite the fact that our physically-based model cannot be directly applied in the case of catchments with an unknown mean flow, the comparison of its results for gauged catchments with estimates from the FEH method allows us to better understand why the latter approach is inaccurate in the case of small catchments. There are two main reasons for this, related to estimating the two main factors in equation \eqref{eq:QMED_new}, namely the mean flow ($\bar{q}$) and the base flow index (BFI).

\begin{figure}
    \centering
    \includegraphics{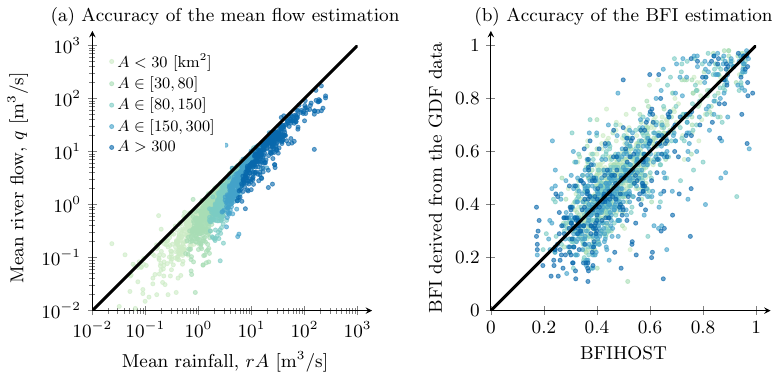}
    \caption{Two sources of inaccuracy of the QMED estimation method. Each point represents one NRFA catchment, and the colour represents their area. Fig. (a) presents the difference between the precipitation rate and mean river flow, related to evapotranspiration and other factors affecting water balance. Fig. (b) presents the difference between BFI estimated from HOST classification and directly from gauged daily flow data.}
    \label{fig:sources_of_inaccuracy}
\end{figure}

Firstly, as demonstrated in \cref{sec:statistical_methods}, the currently used FEH catchment descriptors are not sufficient to describe the mean river flow $\bar{q}$. Factors such as artificial groundwater extraction, vegetation abundance, and mining activities significantly affect the amount of precipitation reaching the river. The effect is most notable in small catchments, as illustrated in \cref{fig:sources_of_inaccuracy}a. Note that the difference between precipitation and river flow has a much higher variance for small catchments with low mean river flow values, since they are more prone to local human activities. In order to obtain reliable predictions of mean flow, either all significant factors affecting the water balance should be taken into account, or the mean flow should be measured directly.

The second source of inaccuracy is the method of estimating the base flow index based on the HOST soil classification (BFIHOST). As shown in \cref{fig:sources_of_inaccuracy}b, there is significant uncertainty related to this estimation method for catchments of all sizes. As we have shown, this inaccuracy has a significant impact on the predictions of both compared models. Therefore, we suggest developing a better measure for BFI by including factors other than soil type alone. According to our physical benchmark model, BFI should depend on factors such as mean precipitation rate, hillslope width, soil depth, and elevation gradient.

Further work on these two aspects could reduce the inaccuracy of peak flow prediction from $\text{SD}(e)=0.27$ down to $\text{SD}(e)=0.12$ for the $10\%$ smallest catchments in the NRFA dataset (\emph{i.e.}, catchments with an area below $31.5;\mathrm{km}^2$). The remaining inaccuracies are related to the mechanisms of generating peak flow, which were not captured by our simple model \eqref{eq:QMED_new}.

\section{Conclusions}

In this paper, we have derived a simple physical model for peak flow formation based on our earlier study of coupled surface-subsurface flows at the catchment scale. The model has the same structure as commonly used statistical models and is equally simple to apply, but it is based on a clear set of theoretical assumptions and does not require calibration.

The physical model offers two important benefits over data-driven statistical methods. Firstly, it provides a clear physical interpretation behind the formula, allowing for a better understanding of how flow patterns may change, such as in different seasons or due to climate change, without the need for regular recalibration. Secondly, due to the lack of calibration, the model is independent of data availability, making it equally accurate in all types of situations, including those underrepresented in standard datasets (\emph{e.g.} small catchments). This also reduces the risk of overfitting.

As demonstrated, the model exhibits high precision in predicting monthly peak river flow across all UK catchments from the National River Flow Archive, and annual peak flows after applying the approximation \eqref{eq:QMED_model} to the QMED formulation. The model outperformed the QMED estimation method from the Flood Estimation Handbook, resulting in lower residuals (and therefore higher $R^2$ values), and significantly fewer outliers. Importantly, the model maintained its high precision even for small catchments, which is crucial for predicting peak flows in ungauged catchments, a challenging problem in hydrology~\citep{reed1999flood}.

Although the model cannot be directly applied to ungauged catchments as it requires an estimate of the mean river flow, it sheds light on the limitations of the FEH method. We have shown that the problematic aspect is not the understanding of peak flow formation, but rather the inability of current FEH catchment descriptors to accurately predict mean rainfall (and consequently the size of the seepage zone). Additionally, we found that the method of estimating the base flow index based on HOST soil classification introduces inaccuracies to predicted peak flows. Further investigation of these issues is recommended to develop more robust statistical models for peak flow estimation in ungauged catchments.

The high performance of this simple physical model demonstrates the significant potential of physically-based benchmark models in improving existing flood estimation methods. Our increasing understanding of fundamental hydrologic processes at the catchment scale can greatly enhance exploratory modelling beyond what can be learned from data exploration alone. We agree with~\cite{kirchner2006getting} that in order to advance the science of hydrology, we not only have to construct models consistent with the available data, but models that give correct predictions for well-understood reasons. Only then can we rely on the model's performance when applied to situations not captured or underrepresented in the available data, which we believe is crucial to solving such challenges in hydrology as estimating peak flows in small ungauged catchments or predicting the effect of climate change on flood patterns.

Therefore, we strongly encourage researchers in the field of hydrology to take advantage of advancements in physical modelling and benchmarking to better understand the limitations of currently used hydrologic models and to develop more theoretically justified and robust models in the future.

\mbox{}\par
\noindent \textbf{Acknowledgements.} We thank Sean Longfield (Environmental Agency) for useful discussions, and for motivating this work via the 7th Integrative Think Tank hosted by the Statistical and Applied Mathematics CDT at Bath (SAMBa). We also thank Thomas Kjeldsen, Tristan Pryer, Keith Beven and Simon Dadson for insightful discussions. Piotr Morawiecki is supported by a scholarship from the EPSRC Centre for Doctoral Training in Statistical Applied Mathematics at Bath (SAMBa), under the project EP/S022945/1.

\bibliographystyle{plainnat}
\bibliography{bibliography}

\newpage
\appendix

\section{Precision of peak flow estimation for each month}
\label{app:monthly_summary}

\begin{figure}
    \centering
    \includegraphics{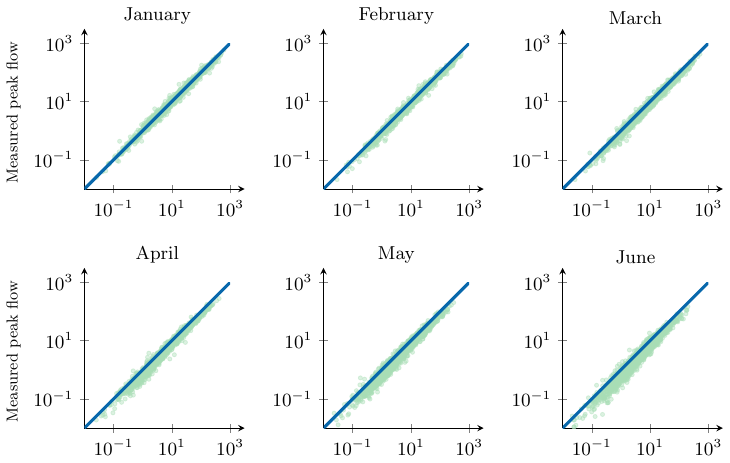}
    \includegraphics{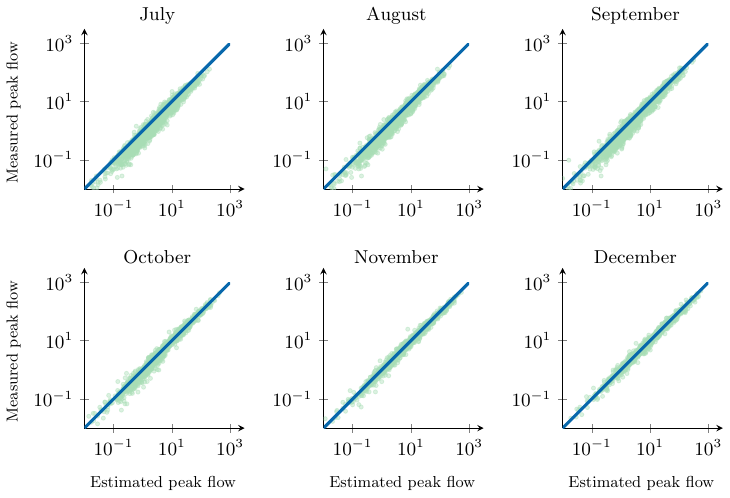}
    \caption{Reproduction of \cref{fig:our_model_fit} for all months.}
    \label{fig:seasons}
\end{figure}

In \cref{sec:monthly_peak_flow_est}, we analysed the accuracy of equation \eqref{eq:QMED_new} in predicting the median of maximum flow in January. We conducted a similar analysis for other months and summarized the results in \cref{fig:seasons} and \cref{tab:monthly_summary}.

\begin{table}
    \centering
    \begin{tabular}{lccc}
        Month & $\text{MEAN}(R)$ & $\text{SD}(R)$ & $R^2$ \\
        \hline
        January & $-0.016$ & $0.19$ & $0.036$ \\
        January & $-0.016$ & $0.19$ & $0.036$ \\
        February & $-0.15$ & $0.19$ & $0.059$ \\
        March & $-0.21$ & $0.20$ & $0.086$ \\
        April & $-0.31$ & $0.25$ & $0.16$ \\
        May & $-0.38$ & $0.30$ & $0.23$ \\
        June & $-0.45$ & $0.35$ & $0.33$ \\
        July & $-0.43$ & $0.40$ & $0.34$ \\
        August & $-0.33$ & $0.38$ & $0.25$ \\
        September & $-0.29$ & $0.41$ & $0.25$ \\
        October & $-0.10$ & $0.31$ & $0.10$ \\
        November & $-0.071$ & $0.22$ & $0.056$ \\
        December & $-0.033$ & $0.20$ & $0.042$
    \end{tabular}
    \caption{Summary of residual values for each month. Residuals were computed using~\eqref{eq:residuals}.}
    \label{tab:monthly_summary}
\end{table}

\end{document}

%% file: FINALFIG/saturated_zone.pdf_tex
\begingroup%
  \makeatletter%
  \providecommand\color[2][]{%
    \errmessage{(Inkscape) Color is used for the text in Inkscape, but the package 'color.sty' is not loaded}%
    \renewcommand\color[2][]{}%
  }%
  \providecommand\transparent[1]{%
    \errmessage{(Inkscape) Transparency is used (non-zero) for the text in Inkscape, but the package 'transparent.sty' is not loaded}%
    \renewcommand\transparent[1]{}%
  }%
  \providecommand\rotatebox[2]{#2}%
  \newcommand*\fsize{\dimexpr\f@size pt\relax}%
  \newcommand*\lineheight[1]{\fontsize{\fsize}{#1\fsize}\selectfont}%
  \ifx\svgwidth\undefined%
    \setlength{\unitlength}{285.50619269bp}%
    \ifx\svgscale\undefined%
      \relax%
    \else%
      \setlength{\unitlength}{\unitlength * \real{\svgscale}}%
    \fi%
  \else%
    \setlength{\unitlength}{\svgwidth}%
  \fi%
  \global\let\svgwidth\undefined%
  \global\let\svgscale\undefined%
  \makeatother%
  \begin{picture}(1,0.38051888)%
    \lineheight{1}%
    \setlength\tabcolsep{0pt}%
    \put(0.60920177,0.00567146){\color[rgb]{0,0,0}\makebox(0,0)[lt]{\lineheight{1.25}\smash{\begin{tabular}[t]{l}Realistic catchment\end{tabular}}}}%
    \put(0,0){\includegraphics[width=\unitlength,page=1]{saturated_zone.pdf}}%
    \put(0.06707516,0.00506751){\color[rgb]{0,0,0}\makebox(0,0)[lt]{\lineheight{1.25}\smash{\begin{tabular}[t]{l}Simplified catchment\end{tabular}}}}%
    \put(0,0){\includegraphics[width=\unitlength,page=2]{saturated_zone.pdf}}%
    \put(0.15559933,0.35836388){\color[rgb]{0,0,0}\makebox(0,0)[lt]{\lineheight{1.25}\smash{\begin{tabular}[t]{l}Saturated zone\end{tabular}}}}%
    \put(0.46144919,0.35859446){\color[rgb]{0,0,0}\makebox(0,0)[lt]{\lineheight{1.25}\smash{\begin{tabular}[t]{l}Unsaturated zone\end{tabular}}}}%
    \put(0.79714417,0.35839221){\color[rgb]{0,0,0}\makebox(0,0)[lt]{\lineheight{1.25}\smash{\begin{tabular}[t]{l}Channel\end{tabular}}}}%
    \put(0.20509586,0.18310328){\color[rgb]{0,0,0}\makebox(0,0)[lt]{\lineheight{1.25}\smash{\begin{tabular}[t]{l}$A_s$\end{tabular}}}}%
    \put(0,0){\includegraphics[width=\unitlength,page=3]{saturated_zone.pdf}}%
    \put(0.78121618,0.23538089){\color[rgb]{0,0,0}\makebox(0,0)[lt]{\lineheight{1.25}\smash{\begin{tabular}[t]{l}$A_s$\end{tabular}}}}%
    \put(0,0){\includegraphics[width=\unitlength,page=4]{saturated_zone.pdf}}%
  \end{picture}%
\endgroup%

%% file: TIKZ/res_vs_descriptors.tex
\begin{tikzpicture}[font=\small]

\newcommand{\resplot}[8] {

\definecolor{greenPoint}{RGB}{168,221,181}
\definecolor{greenLine}{RGB}{69,175,95}
\definecolor{greenFill}{RGB}{69,175,95}

\definecolor{bluePoint}{RGB}{67,162,202}
\definecolor{blueLine}{RGB}{8,104,172}
\definecolor{blueFill}{RGB}{8,104,172}

\begin{axis}[
    mark size=2.9pt,
    width=0.45\linewidth,
    height=0.35\linewidth,
    at={(#1,#2)},
    xlabel={#3},
    ylabel={Log residual},
    xmin=#5,xmax=#6,
    ymin=-1,ymax=1,
    grid=both,
    grid style={draw=gray!25},
    clip marker paths=true,
    xmode = #7,
    axis lines = left,
    axis line style={shorten >=-7pt},
    legend style={at={(-0.4,-0.4)},anchor=north},
    legend columns=-1,
]

\addplot [color=bluePoint, mark=x, only marks, mark size=1pt, opacity=0.8] table [x="x", y="res2", col sep=comma] {DATA/res_vs_descriptors/#4.dat};

\addplot [color=greenPoint, mark=+, only marks, mark size=1pt, opacity=0.8] table [x="x", y="res1", col sep=comma] {DATA/res_vs_descriptors/#4.dat};

\addplot [blueLine, line width=1pt] table [x=x, y=y, col sep=comma] {DATA/res_vs_descriptors/#4_line_2.dat};
\addplot [name path=f, blueFill, line width=0.5pt] table [x=x, y=ymin, col sep=comma] {DATA/res_vs_descriptors/#4_line_2.dat};
\addplot [name path=g, blueFill, line width=0.5pt] table [x=x, y=ymax, col sep=comma] {DATA/res_vs_descriptors/#4_line_2.dat};
\addplot[blueFill, opacity=0.4] fill between[of=f and g];

\addplot [greenLine, line width=1pt] table [x=x, y=y, col sep=comma] {DATA/res_vs_descriptors/#4_line_1.dat};
\addplot [name path=f, greenFill, line width=0.5pt] table [x=x, y=ymin, col sep=comma] {DATA/res_vs_descriptors/#4_line_1.dat};
\addplot [name path=g, greenFill, line width=0.5pt] table [x=x, y=ymax, col sep=comma] {DATA/res_vs_descriptors/#4_line_1.dat};
\addplot[greenFill, opacity=0.4] fill between[of=f and g];

\ifthenelse{#8=1}{\legend{FEH method~~~~, Physical benchmark}}{}

\end{axis}
}

\resplot{0}{0}{AREA [$\mathrm{km^2}$]}{catchment.area.x}{1}{1e4}{log}{0};
\resplot{0.5\linewidth}{0}{FARL}{farl}{0.6}{1}{normal}{0};

\resplot{0}{-0.35\linewidth}{BFI}{gdf.base.flow.index}{0}{1}{normal}{0};
\resplot{0.5\linewidth}{-0.35\linewidth}{LCM2017 urban}{lcm2007.urban}{0}{1}{normal}{1};

\end{tikzpicture}